# Tau Herculids

## Tau Herculids 2022: Rate, number density, population index and geometrical effects from visual data

*Jürgen Rendtel [1,2] and Rainer Arlt [1]*

We analysed visual observation data of the Tau Herculids collected between 2022 May 28 and June 1. The population index $r$ is 2.4 for entire sample. For the pre-peak period we find $r = 2.57 \pm 0.23$, the peak period yields $r = 2.38 \pm 0.06$. The ZHR maximum of the 1995 ejecta from comet 73P/Schwassmann-Wachnann3 (SW3) is found at $\lambda_\odot = 69°\!.450$, i.e. May 31, $05^{\mathrm{h}}04^{\mathrm{m}}$ UT ($\pm 5$ min) with a ZHR $= 55 \pm 7$, corresponding with a spatial number density ND $\approx 380 \times 10^{-9}\mathrm{km}^{-3}$ for meteoroids larger than 10 mg. An earlier maximum occurred at $\lambda_\odot = 69°\!.207$, i.e. May 30, centred at $23^{\mathrm{h}}$ UT with a ZHR $= 18 \pm 3$ and is tentatively associated with SW3-ejecta from 1892. Effects of the radiant shift due to the large zenith attraction of about 10° for the radiant close to the horizon are discussed.

## 1 Introduction

The encounter with the fresh debris of comet 73P/Schwassmann-Wachmann 3 (short SW3) on 2022 May 31 (described in detail e.g. in Rao, 2021) caused many observational efforts worldwide to observe the event under favourable circumstances. The very first calculations including the precise prediction of the encounter time was published by Lüthen et al. (2001). The first as well as later predictions of the major peak due to meteoroids released from SW3 after the breakup in 1995 agreed very well in their timing. A summary of recent predictions is given e.g. in the IMO 2022 Meteor Shower Calendar (Rendtel, 2021). The (predicted) times are:

May 31, 04:55 UT ($\lambda_\odot = 69°\!.44$; min. dist. $+0.0004\,\mathrm{au}$; Jenniskens 2006, quoting Lüthen et al. 2001),

May 31, 05:17 UT ($\lambda_\odot = 69°\!.459$; $-0.00214\,\mathrm{au}$; Jenniskens 2006),

May 31, 05:04 UT ($\lambda_\odot = 69°\!.451$; $-0.00041\,\mathrm{au}$; Sato 2021).

Sato commented "the density of the trail is estimated to be low because of the large ejection velocity. However, we may be able to see a meteor storm [...] because a lot of dust is expected due to the breakup". Another very late calculation by Vaubaillon[a] yielded a peak centered at $05^{\mathrm{h}}01^{\mathrm{m}}$ UT. Additionally, the Earth was expected to encounter SW3-dust mainly ejected in 1892 and 1897 (Wiegert et al., 2005). The diagram shown in this paper indicates an encounter of the 1897 meteoroids near $16^{\mathrm{h}}$ UT and of the more widely scattered 1892 meteoroids close to $02^{\mathrm{h}}$ UT.

Therefore all observations between May 30, about $16^{\mathrm{h}}$ UT and May 31, about $07^{\mathrm{h}}$ UT were of essential interest. Some of the questions which possibly can be answered from analyses of visual data are:

- times of ZHR maxima, especially the main peak,
- strength of the ZHR maxima,
- difference in the magnitude distributions (population index $r$),
- correlation between ZHR and $r$,
- number density in the different regions of the stream.

It is not possible to observe the entire period of interest from one location on Earth. At the end of May, the number of dark hours in the northern hemisphere is rather short and observers north of about 55° N cannot observe at all. Locations in wide parts of North America were favourable to observe the main peak. European locations allowed to follow part of the early activity. Depending on the latitude, the dark window was quite short. From the Canary Islands it was possible to observe for 8 hours and thus to see a part of the early activity as well as the time until the main peak. We found that 05:10 UT is the latest moment for useful data when the western sky is still reasonably dark while the the eastern sky has already bright twilight. Additionally, the radiant elevation decreases a lot towards the morning. Hence all effects around the zenith attraction za (see Koschack et al., 2022) and zenith coefficient $\gamma$ (see Bellot Rubio, 1995) are of great importance. Our own short campaign on Tenerife from May 28 to June 1 with four clear nights in a row was successful and contributed data for the maximum night as well as from the neighbouring nights for calibration purposes. This allows us to have a closer look at the radiant change due to gravitation effects but not on the zenith coefficient question.

## 2 Visual observations worldwide

The IMO's VMDB received reports from 45 visual observers covering the period from May 25 to June 3 (with the vast majority of reports from the maximum

[1]Leibniz-Institut f. Astrophysik Potsdam, An der Sternwarte 16, 14482 Potsdam, Germany. Email: jrendtel@web.de
[2]International Meteor Organization, Eschenweg 16, 14476 Potsdam, Germany.

night May 30/31), including data of 1661 shower meteors. Here we list the observers and their observing region as well as the number of session reports submitted:

Mark Adams (USA, 3 sessions); Daniel Alcázar (Spain, 1); Rainer Arlt (Spain, 4); Orlando Benítez Sánchez (Spain, 1); Tim Cooper (South Africa, 1); Howard Edin (USA, 1); Christoph Gerber (Germany, 1); Robert Harris (USA,1 ); Jan Hattenbach (Spain, 1); Carl Hergenrother (USA, 1); Glenn Hughes (Australia, 9); Javor Kac (USA, 2); André Knöfel (Germany, 1); Pete Kozich (USA, 1); Jens Lacorne (France, 1); Anna Levin (Israel, 1); Michael Linnolt (USA, 1); Robert Lunsford (USA, 3); Oleksandr Maidyk (Ukraine, 1); Oscar Martin Mesonero (Spain, 1); Pierre Martin (Canada, 2); Marco Micheli (Italy, 1); Russell Milton (USA, 2); Koen Miskotte (France, 6); Sirko Molau (USA, 1); Edward Murphy (USA, 1); Basil Nikolau (USA, 1); Artyom Novichonok (Russia, 3); Francisco Ocaña González (USA, 2); Sasha Prokofyev (Cyprus, 4); Ina Rendtel (Germany, 3); Jürgen Rendtel (Germany, Spain, 6); Terrence Ross (USA, 3); Ivan Sergey (Belarus, 4); Wesley Stone (USA, 1); Fengwu Sun (USA, 1); Hanjie Tan (Czech Republic, 1); Austin Uhler (USA, 1); Michel Vandeputte (Belgium, 1); Alan Webb (USA, 2); Thomas Weiland (USA, 1); Frank Wächter (Germany, 1); Sabine Wächter (Germany, 2); Quanzhi Ye (USA, 1)

The nights May 28/29 to May 31/June 1 are well covered, and we have a continuous data series from May 30, $21^h$ UT, to May 31, close to $12^h$ UT. This allows us to analyse the points raised in the Introduction. A first look into the data, e.g. as provided by the IMO live graph, shows a ZHR profile with a main peak near $05^h15^m$ UT. Additionally, we find an earlier and much weaker maximum close to May 30, $23^h$ UT. It may be tentatively associated with the meteoroids released from the comet around the perihelia in 1892 and 1897.

In order to obtain complete information, we first analyse the magnitude data before dealing with the ZHR and spatial number density. Due to the very low entry velocity, the visual meteors represent a mass range which considerably differs from (most of) the known meteor showers. The consequences are described below.

## 3 Population index $r$

The ZHR calculation requires the knowledge of the population index $r$ to correct for the standard conditions (see, in detail, Koschack et al., 2022). The available observing reports cover the period between May 30, $21^h$ UT, and May 31, $12^h$ UT quite well. The geographical distribution of the observers between Eastern Europe and Western North America as listed above ensures that we have magnitude (and rate) data with high radiant position available for the entire period.

First, we calculated a general value of the population index for the shower including both the main peak of fresh meteoroids and the early activity period caused by meteoroids released about a hundred years earlier. The general average is $r = 2.40 \pm 0.06$ centred at $\lambda_\odot = 69°\!.407$ and is based on magnitude data of 1521 TAH meteors (245 intervals with magnitude distributions).

Next, we checked whether there is a difference between the main activity period and the period before. The respective values are: old meteoroids $r = 2.57 \pm 0.23$ centred at $\lambda_\odot = 69°\!.198$ and fresh meteoroids $r = 2.38 \pm 0.06$ centred at $\lambda_\odot = 69°\!.455$.

It seems the values further to the edges are slightly higher. We find $r = 2.81 \pm 0.67$ at $\lambda_\odot = 68°\!.36$ (only based on magnitude data of 35 TAH), and $r = 2.44 \pm 0.92$ at $\lambda_\odot = 70°\!.15$ (17 shower meteors).

The two values for old and fresh indicate that there is a small difference in the meteoroid size distribution between the fresh material and the meteoroids ejected earlier. However, the difference is not remarkably large.

One of the main questions for the 2022 activity was, whether meteoroids were able to reach the Earth because the "standard ejection conditions" would have hardly brought meteoroids released as a consequence of the 1995 comet breakup to Earth encounters (see the remarks in the Introduction and at the IMCCE website quoted above). A higher ejection speed was required, but it seemed open what size the encountering meteoroids would have. A question, which was of great importance because of the low velocity of the TAH meteors.

The observed meteor magnitude range (Table 1) between $-3$ and $+5$ mag translates into a mass range between 170 g (!) and 0.06 g. Meteoroids of the same mass range would appear as meteors of $-6.1$ and $+1.6$ mag, respectively, when entering the atmosphere at 35 km/s (Geminids), and much brighter if apearing as Perseids. This means, that the bright Tau Herculids (of $-2$ or $-3$ mag) which we saw particularly around the peak (Table 2) were quite large meteoroids which are not frequent in other showers. The reported magnitude distributions include rather few $+6$ magnitude TAH meteors in the magnitude data. This is a bit surprising because the original expectation was that we may see a shower rich in faint meteors due to the low velocity. So the visual data raise the opposite question, whether the comet mainly released larger meteoroids (in the recent breakup ejection as well as in the older material) – or whether this is an observers' bias. The latter seems unlikely, as the apparent lack of $+6$ TAH meteors is found throughout the entire activity period.

*Table 1* – Magnitudes of the observed Tau Herculid meteors in different periods. The first line gives the total of all TAH meteors from reported between May 24 to June 4; the subsequent lines (labelled 'Max.') give details for periods of the maximum night May 30/31. There is one $-6$ TAH meteor which was seen by two observers at the same site. Two session reports summarised the magnitude data over 2.5 and 3.0 hours, respectively, and are not considered in the separate pre- / post-peak distributions.

| Sample | $-6$ | $-5$ | $-4$ | $-3$ | $-2$ | $-1$ | 0 | $+1$ | $+2$ | $+3$ | $+4$ | $+5$ | $+6$ |
|---|---|---|---|---|---|---|---|---|---|---|---|---|---|
| All TAH meteors | 2 | 0 | 2 | 8 | 12.5 | 33 | 84 | 204 | 277.5 | 412.5 | 448 | 298.5 | 30 |
| Max. $19^h$–$02^h$UT | 0 | 0 | 0 | 0 | 1.5 | 2.5 | 8.5 | 24.5 | 26.5 | 41 | 74.5 | 38.5 | 10.5 |
| Max. $02^h$–$05^h$UT | 0 | 0 | 1 | 5 | 2 | 16.5 | 34 | 73.5 | 115.5 | 195 | 184 | 131 | 9.5 |
| Max. $05^h$–$11^h$UT | 2 | 0 | 1 | 3 | 7 | 11 | 37.5 | 92 | 115.5 | 158 | 156 | 110 | 8 |

Table 2 – Appearance of bright Tau Herculids $TAH_B$ (here: $-2$ mag and brighter) in different sections of the activity profile close to the maximum on 2022 May 30/31. The time given here is the total effective observing time of all observers contributing to the sample. The $-6$ mag TAH meteor observed near $06^h30^m$ UT is one of the "4" in the line 0600–0700 UT although referring to one single fireball (reducing the $TAH_B$ to 0.30 if counted as just one bright TAH).

| Period (UT) | #$TAH_B$ | Eff. obs. time (total, hrs) | $TAH_B$/hr |
|---|---|---|---|
| < 11 | 0 | | 0 |
| 1940–0200 | 1.5 | 27.2 | 0.055 |
| 0200–0300 | 0 | 2.5 | 0 |
| 0300–0400 | 3 | 12.0 | 0.25 |
| 0400–0500 | 4 | 15.1 | 0.27 |
| 0500–0520 | 5 | 4.9 | 1.02 |
| 0520–0600 | 0 | 10.7 | 0 |
| 0600–0700 | 4 | 10.1 | 0.40 |
| 0700–0800 | 1 | 4.7 | 0.22 |
| > 08 | 0 | 7.8 | 0 |

The amount of magnitude data allows us to try looking for details within the two "activity periods", i.e. for structures in the stream. We adjusted the binning lengths for the $r$-calculation throughout the entire period. For the interval $69°.13$ to $69°.35$ (about $21^h$ UT to $02^h30^m$ UT) we used bins of $0°.08$ shifted by $0°.04$ (giving a temporal resolution of 1 hour). The large number of data around the main peak allowed us to set the bin length $0°.04$ shifted by $0°.02$ (30 minute resolution). The result is shown in Figure 1.

Encouraged by this surprisingly smooth profile, we tried even shorter bins, being aware that the error margins and uncertainties become much larger. Nevertheless we think that the profile with 10 minute bins ($0°.007$) has some information. An inspection of the magnitudes of bright shower meteors revealed that there was a kind of stop at $05^h15^m$ UT or immediately after that. Before this, a significant number of $-2$ to $-3$ mag meteors was reported, but almost none in the period after that. However, a change of the population index from 2.26 to 2.50 from one 10-minute interval to the next occurs only after $05^h30^m$ UT. Like all the variations we see in the profile shown in Figure 2, the error margins indicate that cannot draw conclusions from any feature – even if we find confirmation by other data series.

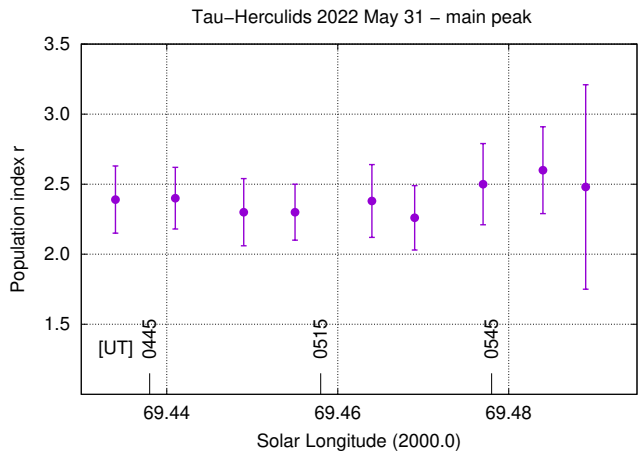

Figure 2 – Attempt to derive a population index $r$ profile of the Tau Herculids with higher temporal resolution (10 minutes) during the main peak period. Details are given in the text. The variations are rather small and we cannot find features in the profile.

## 4 ZHR profile

For the calculation of the ZHR values we applied the $r$-profile shown in Figure 1. A large fraction of the observers had good observing conditions (limiting magnitude $+6$ or better) so that a difference of say 0.1 of the population index does not affect the resulting ZHRs too much. The result of the ZHR calculation is shown in Figure 3. Like in the case of the population index, the smaller sample for intervals when the old TAH meteoroids occurred, we also used longer bins for the ZHR calculation.

There are two obvious features in the ZHR profile: a sharp and pronounced peak with a ZHR $= 55 \pm 7$ at $05^h05^m$ UT ($\pm 5$ min) and a rather broad maximum with a ZHR $= 18 \pm 3$ centred at $23^h$ UT. The main peak has a skew shape. The ZHR reaches half the peak value (25) near $02^h40^m$ UT (duration 2.5 hours) and the descend to the same ZHR happens close to $06^h30^m$ UT (duration 1.3 hours). From our data we cannot see whether the longer ascend is a characteristic of the distribution of the fresh meteoroids or a superposition with probable older material. This may perhaps be distinguished from orbital data.

The early maximum with a ZHR just below 20 most likely is caused by meteoroids from the old particle releases. Considering the diagram shown in Wiegert et al. (2005), the broader scattered 1892 meteoroids are more likely the cause for the maximum we find around $23^h$ UT. We may estimate a width of this maximum of about 3 hours, but as the end seems to overlap with the starting ascend to the main peak, this is somewhat uncertain.

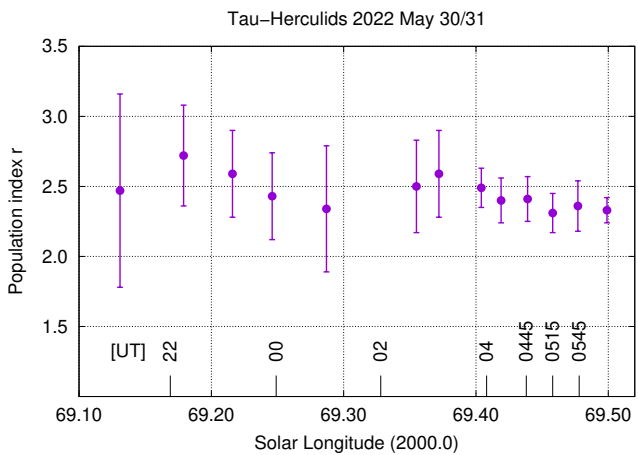

Figure 1 – Population index $r$ of the Tau Herculids during the night 2022 May 30-31 with 1-hour resolution before $02^h$ UT and 30-minute resolution around the peak (see the bin lengths given in the text). The much larger error margins before $02^h$ UT are caused by the smaller sample available for this period of lower activity.

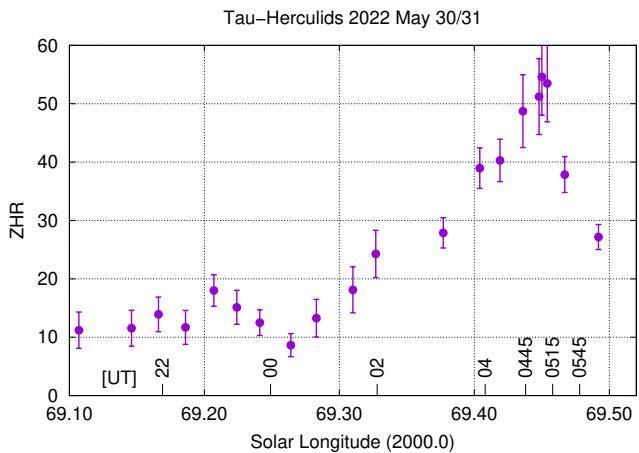

*Figure 3* – ZHR-profile of the Tau Herculids during the night 2022 May 30-31, using the population index profile shown in Figure 1. The peak occurred at $\lambda_\odot = 69°\!.450$, on May 31, $05^h05^m$ UT. An earlier maximum is found at $\lambda_\odot = 69°\!.207$, on May 30, $23^h$ UT.

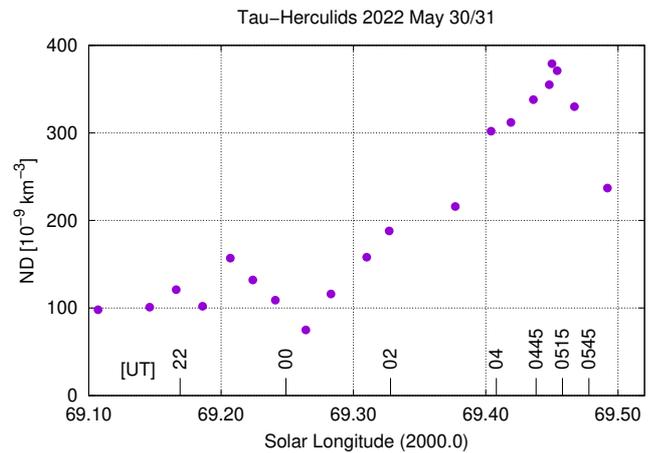

*Figure 4* – Profile of the spatial number density of the Tau Herculids during the night 2022 May 30-31, based on the ZHR profile shown above and the population index as shown in Figure 1.

## 5 Number density

As pointed out above, the TAH meteor brighnesses up to $-2$ to $-3$ magnitude (very few meteors even brighter than that!) and the low atmospheric entry velocity are equivalent to rather large meteoroid masses. The number density, which is the better suited quantity to compare meteoroid streams than the ZHR (which describes the appearance of the shower in the observer's sky), is remarkably high (Figure 4).

The Geminids with a ZHR of about 150 are caused by a meteoroid stream with a spatial number density $ND \approx 200 \times 10^{-9} \text{km}^{-3}$ for meteoroids larger than 1 mg; the corresponding value for the Tau Herculid peak is about $380 \times 10^{-9} \text{km}^{-3}$ – for meteoroids larger than 10 mg (we do not have fainter meteors / smaller meteoroids in our sample). So the TAH shower is about twice as dense as the Geminids. Not to imagine the impression if the Earth would enter such a stream at Perseid speed.

## 6 Some thoughts on the zenith correction

The correction of observed meteor rates to a situation in which the shower radiant is in the zenith has been a long-standing issue. For meteoroids on parallel trajectories hitting a plane "detector", it is $\sin^{-1} h_R$, where $h_R$ is the elevation of the radiant above that plane. Various attempts have included the curvature of the Earth's surface, which in principle allows for meteors becoming visible if the radiant is slightly below the horizon. The issue was treated by, e.g., Kresák (1954) and Richardson (1999). Both authors appear to take the apparent radiant elevation as the relevant quantity to compute the zenith correction factor.

The apparent radiant is the one resulting from the vector addition of the orbital motions of the stream particles and the Earth, plus the shift of the radiant towards the zenith due to the gravitational attraction of the particles by the Earth, before they light up as meteors in the atmosphere. This shift is individual to each observer location and time and needs to be computed for each observing period. (There is also another velocity vector of the observer which needs to be added to the radiant direction. It results from the rotation of the Earth and is called diurnal aberration, but has very small effects on the corrections of visual observations.)

Whether the zenith corrections needs to be based on the apparent radiant is not entirely obvious. The question comes down to asking: does a detector on Earth record the same number of particles from a shower with apparent radiant elevation $h_A$, i.e. *after* gravitational attraction, as a detector in space without gravity and therefore entirely geometrical radiant angle $h_R$ which is equal to $h_A$? In other words, does the zenith attraction simply change the direction of the particle flux vector without affecting the absolute flux?

As demonstrated by Gural (2001), the radiant of particles ideally moving in parallel at infinity (radiant an ideal point without gravitational attraction) becomes an area of 4–5 degrees at geocentric velocities like the ones of the tau Herculids. The zenith-attracted radiant locations depend on the locations of particles in three-dimensional space and it may well be that their flux is lower than that of a hypothetical shower unaffected by gravity whose radiant location is equal to the mean zenith-attracted radiant of the real shower. The meteor simulation code by Gural (2002) – employed to e.g. the Leonid shower by Molau et al. (2002) – would be the ideal tool for exploring this question.

To evaluate the effect, we compared a few individual interval data of the present Tau Herculid data of May 31. The zenith distance of the observed radiant $z_O$ affected by the zenith attraction is calculated by $z_O = \frac{z_t}{2} + \arcsin\left(\frac{v_g}{v_\infty} \sin \frac{z_t}{2}\right)$ where $z_t$ is the (geometrical, undisturbed) zenith distance of the radiant, $v_g$ and $v_\infty$ are the velocities before and after the Earth gravitation (for the full details of radiant corrections see Gural, 2001). In the present case $v_g = 12.36$ km/s and $v_\infty = 16.61$ km/s.

*Table 3* – Effect of the zenith correction on the ZHR on 2022 May 31, between $04^{\rm h}$ and $05^{\rm h}05^{\rm m}$ UT. We averaged the individual ZHR values for observers on locations on the Canary Islands (CI) and on various locations in North America (NA). $h_R$ is the geometric radiant elevation; $h_A$ the apparent elevation due to the Earth's gravitation.

| Period (UT) | Intervals (CI) | Radiant $h_R$ | ZHR (avg.) | Radiant $h_A$ | ZHR(CI) (avg.) | ZHR(NA) (avg.) |
|---|---|---|---|---|---|---|
| 0407–0425 | 3 | 12°–17° | 64 | 26°–28° | 38 | 41 |
| 0437–0505 | 6 | 5°–12° | 76 | 17°–23° | 37 | 52 |

For this purpose we find the data recorded from the Canary Islands (CI) and from locations in North America (NA) very useful. They overlap during the period between $04^{\rm h}$ UT and $05^{\rm h}15^{\rm m}$ UT, when the radiant was low in the western sky as seen from the CI (see Figure 5) and near zenith in NA much further west. So the NA data provide the undisturbed ZHR values, and we compare them with the strongly corrected ones (Table 3). The values may suggest that applying the shifted radiant position is too strong. But this is just for a handful numbers and the scatter is enormous. It just demonstrates that the effect is present and the correction acts in the right direction and reliable order of magnitude.

At this point, we emphasize that the quantitative assessment of the zenith correction at very low geocentric meteoroid velocities and very low radiant positions goes beyond the scope of the present paper.

## 7  Discussion

The visual Tau Herculid observations in 2022 allowed us to document the activity of the encounter in great detail. We find two obvious activity maxima. An early maximum (ZHR = $18 \pm 3$ centred at May 30, $23^{\rm h}$ UT) is probably caused by meteoroids released from SW3 in 1892. The observed maximum is roughly 3 hours before the position which is indicated by the modelling (Wiegert et al., 2005). But the meteoroids of this ejection period seem be be scattered over a large range as compared to the main peak.

Unfortunately, we do not have visual data covering the period around $16^{\rm h}$ UT on May 30, when the Earth may have encountered the 1897 SW3-ejecta (again, as modelled).

The main peak occurred at $\lambda_\odot = 69°450$ corresponding to $05^{\rm h}05^{\rm m}$ UT ($\pm 5$ minutes) with a ZHR = $55 \pm 7$. Its shape is skew with a longer ascend than the subsequent descend. From our data we cannot decide at which moment the fresh particle population dominates the observed sample. This may be distinguishable from orbital data. The population index $r$ is slightly higher before or outside the fresh meteoroid range. The difference in $r$ is not really large: 2.57 vs. 2.38. The population index, however, is in the range of other meteor showers. This was not to be expected because a similar particle size distribution to other showers should have resulted in a shower with mainly faint meteors. A large portion of faint TAH meteors was to some extent anticipated but did not happen. Just the opposite: the fresh ejecta from SW3 seem to be larger than average. Perhaps the short duration from the ejection to the observed encounter kept the large meteoroids which may disintegrate with time. If the original size distribution was similar during the 1892/1897 ejections, the difference in $r$ may give a hint at the disintegration process. Perhaps the 1995 SW3 disintegration was unique as it caused large areas of fresh exposed comet surface and releasing an untypical meteoroid sample.

## 8  Conclusions

### 8.1  Observational data

The encounter with the fresh meteoroids was highly anticipated and gained huge attention. Since it was not clear in advance, what level of activity would occur, observers took a lot of effort to collect data applying all techniques.

### 8.2  Observed ZHR

We find two maxima: a pronounced peak ZHR = $55 \pm 7$ $05^{\rm h}05^{\rm m}$ UT ($\lambda_\odot = 69°450$) lasting 3.8 hours (2.5 hours ascend, 1.3 hours descend). It is caused by the 1995 ejecta from SW3 and occurred closest to Sato's and Vaubaillon's most recent prediction (see Introduction). A broad maximum with a ZHR = $18 \pm 3$ occurred at $23^{\rm h}$ UT ($\lambda_\odot = 69°207$); this maximum is about 3 hours wide. It may be associated with dust ejected at the end of the 19th century, but seems to deviate from the modelled distribution as shown in Wiegert et al. (2005) – it is earlier than the probably centre of the 1892 dust.

### 8.3  Population index and meteoroid masses

The population index $r$ is in the same range as for other meteor showers and therefore much lower than expected in advance. We do not see any peculiar change in $r$ around the main peak of fresh meteoroids, but an indication of more bright (of at least $-2$ mag) shower meteors in the immediate vicinity of the peak. The size distribution of the fresh ejecta as well as those released in 1892/97 are not much differing. However, the sizes of the TAH-meteoroids differ considerably from average size distributions found in annual meteor showers.

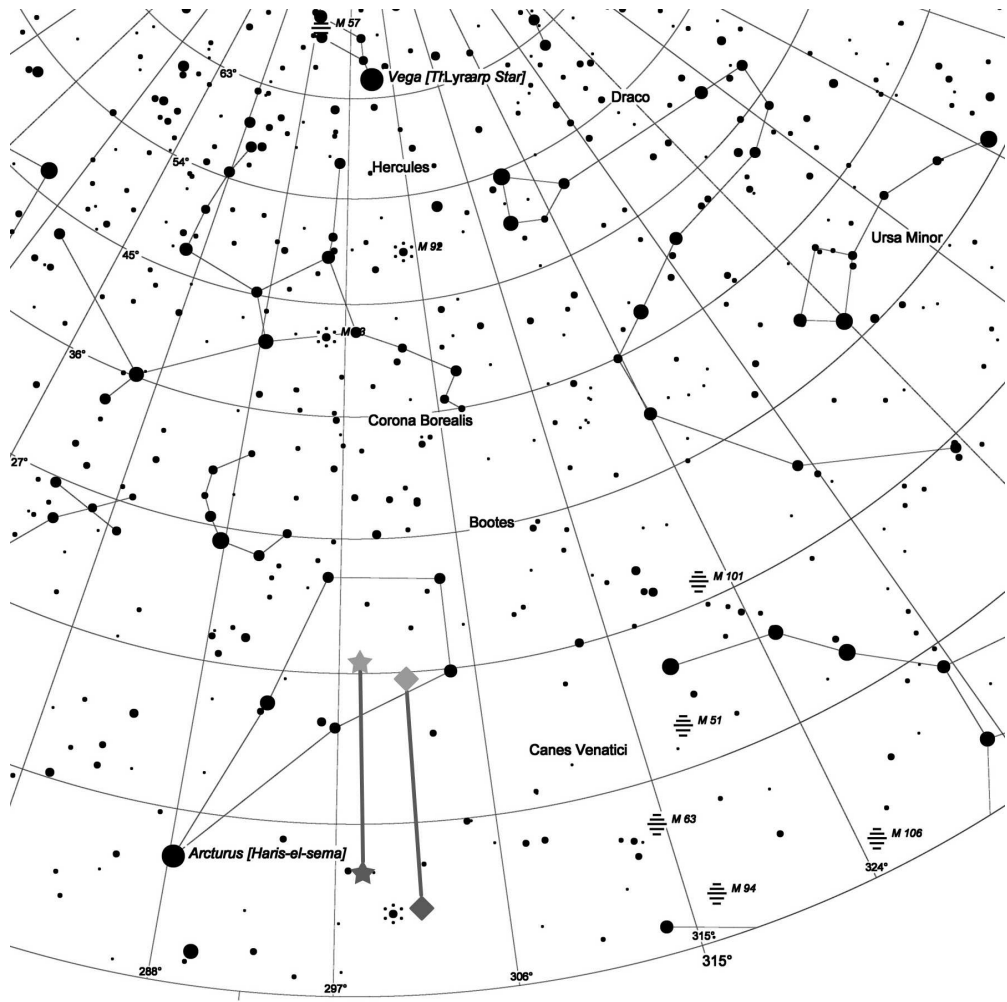

*Figure 5* – Shift of the shower radiant due to the zenith attraction. The setting shown here is for Izaña, Tenerife, on 2022 May 31, 05$^{\rm h}$ UT. The velocities used are $v_g$ = 12.36 km/s (Jenniskens, 2006), $v_{\rm inf}$ = 16.61 km/s (as converted with the Gural (2001) correction). The radiant shift shown as stars refer to the data provided for the 2022 Shower Calendar (Rendtel, 2021) by Sato (personal communication) and Vaubaillon (repeated shortly before the event at https://www.imcce.fr/recherche/campagnes-observations/meteors/2022the), while the the diamonds show the radiant from Lüthen et al. (2001).

### 8.4 Number density

A ZHR of the order of 50 usually does not indicate that the Earth passes a dense stream. Together with the low velocity and the average $r$, we indeed find a dense stream (ND $\approx 380 \times 10^{-9}$km$^{-3}$) for particles of at least 10 mg which has about twice the spatial number density of the Geminid peak (ND $\approx 200 \times 10^{-9}$km$^{-3}$; m $\geq$ 1 mg).

### 8.5 Zenith attraction effects

We briefly discuss in which way the meteoroid trajectories modified by the Earth's gravitation further affect the determination of the ZHR and flux density. A few data obtained under different geometrical conditions indicate that the corrected radiant position needs to be applied for the ZHR (and subsequent flux density) calculation, although we cannot conclusively answer this question in this paper.

### 9 Acknowledgements

We thank all visual observers for submitting their data to the IMO's VMDB. This data flow happened very fast so that we were able to adjust the parameters continuously through the activity period. Therefore, the live graph indeed was a good display of the TAH activity.

### References

Bellot Rubio L. R. (1995). "Effects of a dependence of meteor brightness on the entry angle". *Astronomy & Astrophysics*, **301**, 602.

Gural P. S. (2001). "Fully Correcting for the Spread in Meteor Radiant Positions Due to Gravitational Attraction". *WGN, Journal of the International Meteor Organization*, **29:4**, 134–138.

Gural P. S. (2002). "Meteor Observation Simulation Tool". In Triglav M., Knöfel A., and Trayner C., editors, *Proceedings of the International Meteor Con-*

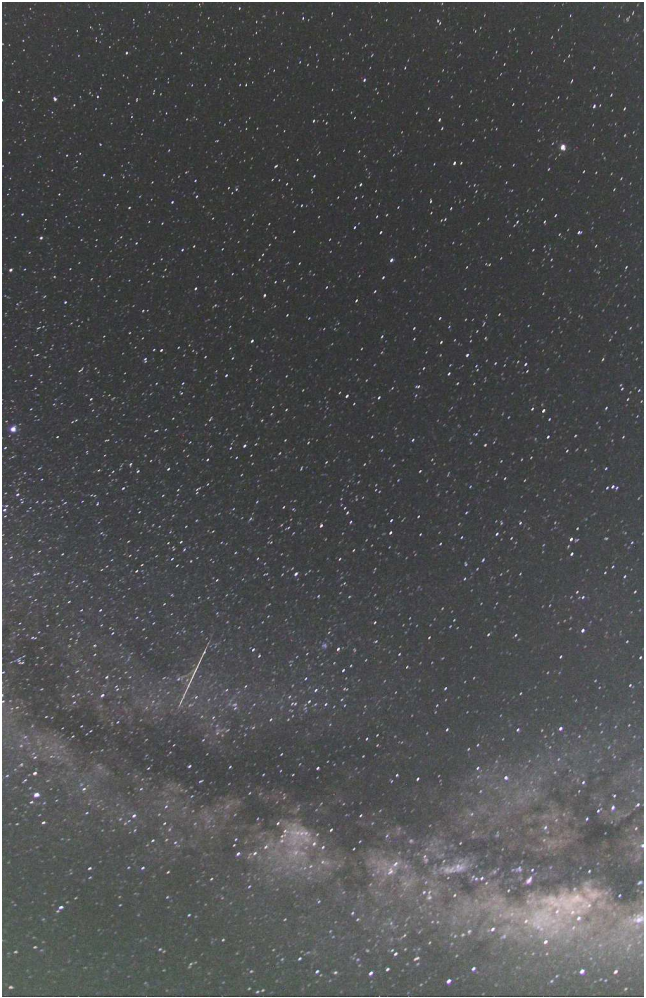

*Figure 6* – Photo of a Tau Herculid meteor taken from Tenerife on 2022 May 31, $01^{\mathrm{h}}11^{\mathrm{m}}$ UT. The backward prolongation of this $-2$ mag meteor already points to a position north of the undisturbed radiant although it appeared about 4 hours before the setting shown in Figure 5. The image was taken with a fish eye lens, $f = 8$ mm, $f/d = 4$, exposed 59 seconds; Canon EOS60Da.

ference, 20th IMC, Cerkno, Slovenia, 2001*. pages 29–35.

Jenniskens P. (2014). "Camelopardalids (IAU#451) from comet 209P/LINEAR". *WGN, Journal of the International Meteor Organization*, **42:3**, 98–105.

Koschack R., Rendtel J., and Richter J. (2022). "Analyses and calculations". In Rendtel J., editor, *Handbook for meteor observers*, page 180.

Kresák L. (1954). "On a Criterion Concerning the Perturbing Action of the Earth on Meteor Streams". *Bulletin of the Astronomical Institutes of Czechoslovakia*, **5**, 45.

Lüthen H., Arlt R., and Jäger M. (2001). "The Disintegrating Comet 73P/Schwassmann-Wachmann 3 and Its Meteors". *WGN, Journal of the International Meteor Organization*, **29**, 15–28.

Molau S., Gural P. S., and Okamura O. (2002). "Comparison of the "American" and the "Asian" 2001 Leonid Meteor Storm". *WGN, Journal of the International Meteor Organization*, **30**, 3–21.

Rao J. (2021). "Will Comet 73P/Schwassman-Wachmann 3 produce a meteor outburst in 2022?". *WGN, Journal of the International Meteor Organization*, **49:1**, 3–14.

Rendtel J., editor (2021). *2022 Meteor Shower Calendar*. International Meteor Organization. IMO_INFO 2-21.

Richardson J. (1999). "A Detailed Analysis of the Geometric Shower Radiant Altitude Correction Factor". *WGN, Journal of the International Meteor Organization*, **27:6**, 308–317.

Wiegert P. A., Brown P. G., Vaubaillon J., and Schijns H. (2005). "The $\tau$ Herculid meteor shower and Comet 73P/Schwassmann-Wachmann 3". *MNRAS*, **361:2**, 638–644.